\documentclass[twocolumn,prb,amsmath,amssymb,floatfix,superscriptaddress,showpacs,longbibliography]{revtex4-1}
\usepackage{graphicx}
\usepackage{dcolumn}
\usepackage{bm}
\usepackage{xspace}
\usepackage{hyperref}
\usepackage{color}
\def\new{\color{black}}
\def\newr{\color{black}}
\UseRawInputEncoding
\def\fgt{Fe$_{3-x}$GeTe$_{2}$\xspace}
\def\fgtn{Fe$_{2.72}$GeTe$_{2}$\xspace}

\begin{document}

\title{Neutron spectroscopy evidence on the dual nature of magnetic excitations in a van der Waals {\new metallic ferromagnet Fe$_{2.72}$GeTe$_{2}$}}
\author{Song~Bao}
\altaffiliation{These authors contributed equally to the work.}
\affiliation{National Laboratory of Solid State Microstructures and Department of Physics, Nanjing University, Nanjing 210093, China}
\author{Wei~Wang}
\altaffiliation{These authors contributed equally to the work.}
\affiliation{School of Science, Nanjing University of Posts and Telecommunications, Nanjing 210023, China}
\author{Yanyan~Shangguan}
\altaffiliation{These authors contributed equally to the work.}
\affiliation{National Laboratory of Solid State Microstructures and Department of Physics, Nanjing University, Nanjing 210093, China}
\author{Zhengwei~Cai}
\altaffiliation{These authors contributed equally to the work.}
\affiliation{National Laboratory of Solid State Microstructures and Department of Physics, Nanjing University, Nanjing 210093, China}
\author{Zhao-Yang~Dong}
\affiliation{Department of Applied Physics, Nanjing University of Science and Technology, Nanjing 210094, China}
\author{Zhentao~Huang}
\author{Wenda~Si}
\affiliation{National Laboratory of Solid State Microstructures and Department of Physics, Nanjing University, Nanjing 210093, China}
\author{Zhen~Ma}
\affiliation{Institute for Advanced Materials, Hubei Normal University, Huangshi 435002, China}
\author{Ryoichi~Kajimoto}
\affiliation{J-PARC Center, Japan Atomic Energy Agency (JAEA), Tokai, Ibaraki 319-1195, Japan}
\author{Kazuhiko~Ikeuchi}
\affiliation{Neutron Science and Technology Center, Comprehensive Research Organization for Science and Society (CROSS), Tokai, Ibaraki 319-1106, Japan}
\author{Shin-ichiro~Yano}
\affiliation{National Synchrotron Radiation Research Center, Hsinchu 30077, Taiwan}
\author{Shun-Li~Yu}
\email{slyu@nju.edu.cn}
\author{Xiangang~Wan}
\author{Jian-Xin~Li}
\email{jxli@nju.edu.cn}
\author{Jinsheng~Wen}
\email{jwen@nju.edu.cn}
\affiliation{National Laboratory of Solid State Microstructures and Department of Physics, Nanjing University, Nanjing 210093, China}
\affiliation{Collaborative Innovation Center of Advanced Microstructures, Nanjing University, Nanjing 210093, China}

\begin{abstract}

In the local or itinerant extreme, magnetic excitations can be described by the Heisenberg model which treats electron spins as localized moments, or by the itinerant-electron model where the exchange interaction between electrons leads to unequal numbers of electrons with up and down spins. However, it has been elusive when both local moments and itinerant electrons are present in the intermediate range. Using inelastic neutron scattering, we provide direct spectroscopic evidence on the coexistence of and interplay between local moments and itinerant electrons in a van der Waals metallic ferromagnet {\new Fe$_{2.72}$GeTe$_{2}$}, which can sustain tunable room-temperature ferromagnetism down to the monolayer limit. We find that there exist ferromagnetic spin-wave excitations dispersing from the zone center at low energies resulting from local moments, and a column-like broad continuum at the zone boundary at high energies up to over 100~meV resulting from itinerant electrons. Unlike the two-dimensional crystal structure, the low-energy mode exhibits a three-dimensional nature, and the high-energy mode also has an out-of-plane dependence. Both modes persist well above the Curie temperature of 160~K. Our neutron spectroscopic data reveal that the low-energy spin waves at 100~K are more coherent than those at 4~K, which is evidence of the weakening of the Kondo screening at high temperatures. These results unambiguously demonstrate the coexistence of local moments and itinerant electrons, and the Kondo effect between these two components in {\new Fe$_{2.72}$GeTe$_{2}$}. Such behaviors are generally expected in heavy-fermion systems with heavy $f$ electrons but rarely clearly observed in materials with light $d$ electrons. These findings shed light on the understanding of magnetism in transition-metal compounds.
\end{abstract}

\maketitle
\section{Introduction}

Magnetism has been a long and mysterious issue in condensed matter physics. Research on it can be divided into two main directions with essentially opposite starting points\cite{moriya1979recent}. First is the local-moment picture initiated by Heisenberg, where each electron is localized and thus produces a local spin moment\cite{zp49_619}. The interaction between localized spins is governed primarily by the Heisenberg exchange coupling. The other is the itinerant-electron model proposed by Bloch\cite{1929ZPhy...57..545B}, Slater\cite{PhysRev.49.537}, and Stoner\cite{S_1947}, where electrons are regarded as itinerant Bloch waves and the exchange interaction between electrons splits the electron band and results in unequal amount of up and down spins, thus giving rise to magnetism\cite{Feng2013,PhysRevB.99.014407}. The magnetic excitations then can have the form of Stoner continuum resulting from the scattering between the particle and hole bands, and therefore depend crucially on the electronic band structure. For the past century, the theory of magnetism has been developed extensively along both directions. However, neither one can be applied to explain the magnetic properties universally. There is a particular challenge in describing the physical properties of magnetic transition metals with $d$ electrons, where the mutually opposite degrees of freedom---localized spins and itinerant electrons are both present. In previous decades, plenty of researches were carried out on the spin dynamics in 3$d$-transition metals like Fe (Refs.~\onlinecite{PhysRev.179.417,PhysRevB.11.2624}) and Ni (Refs.~\onlinecite{PhysRev.182.624,PhysRevLett.30.556,PhysRevB.23.198,PhysRevB.36.881}), and weak itinerant ferromagnets\cite{PhysRevB.16.4956,RevModPhys.79.1015}. Initially, it seemed that the spin dynamics in the ferromagnetic phase could be well described under the framework of spin-wave excitations regardless of the itinerancy\cite{doi:10.1143/JPSJ.75.111002,PhysRev.179.417,PhysRevB.11.2624,PhysRev.182.624,PhysRevLett.30.556,PhysRevB.23.198}. However, the persistent spin waves observed well above the Curie temperature $T_{\rm C}$ were unexpected for a conventional Heisenberg ferromagnet\cite{PhysRevLett.51.300,PhysRevLett.57.150,PhysRevLett.95.087207}, where the spin waves should turn into spin diffusive mode above $T_{\rm C}$ (Refs.~\onlinecite{RevModPhys.39.395,PhysRev.177.952,collins1989magnetic}). To account for the magnetism in these materials, there is a third scenario intermediate between the two aforementioned cases, where some $d$-electrons are itinerant, while the others are localized, and there is interplay between the local moments and itinerant electrons which provide the Kondo screening\cite{RevModPhys.87.855}. The situation is then similar to that found in heavy fermion systems with more localized $f$ electrons\cite{RevModPhys.56.755,Si03092010,npjqm6_60}, but is more controversial due to the more itinerant nature of $d$ electrons\cite{RevModPhys.60.209,PhysRevB.16.4956,nc11_3076,xie2021magnetic,
RevModPhys.78.17,xu2009testing,PhysRevLett.102.187206,zhao2009spin,xulocal1,
np8_709,1367-2630-11-2-025021,2011arXiv1103.5073Z,
PhysRevB.83.214519}. {\new This is particularly the case for iron- and copper-based high-temperature superconductors\cite{PhysRevLett.102.187206,zhao2009spin,xulocal1,np8_709,
1367-2630-11-2-025021,2011arXiv1103.5073Z,
PhysRevB.83.214519,JPCM25_445702,spin1review,BSCCO_Keimer,RevModPhys.78.17,
Nature429_534,helbco,xu2009testing}. On the one hand, both the localized and itinerant models have its own merits and drawbacks in explaining some of the experimental data. On the other hand, direct evidence on the coexistence of local moments and itinerant electrons and their interplay is still lacking.}

Reminiscent of this longstanding debate, we study a recently discovered van der Waals (vdM) metallic ferromagnet Fe$_3$GeTe$_2$. The crystal structure of Fe$_3$GeTe$_2$ belongs to the space group $P6_{3}/mm$ (No.~194)\cite{doi:10.1002/ejic.200501020}, as illustrated in Fig.~\ref{fig:characterizations}(a). It consists of Fe$_3$Ge slabs separated by the vdW-bonded Te layers. In each Fe$_3$Ge slab, Fe ions on Fe(1) and Fe(2) sites are bonded to form a noncoplanar hexagonal network as shown in Fig.~\ref{fig:characterizations}(a) and (b). Neutron powder diffraction has shown that Fe$_3$GeTe$_2$ develops a long-range ferromagnetic order below $T_{\rm C}$, with spins aligned along the $c$ axis as illustrated in Fig.~\ref{fig:characterizations}(a)~(Ref.~\onlinecite{doi:10.1021/acs.inorgchem.5b01260}). The $T_{\rm C}$ of Fe$_{3-x}$GeTe$_2$ usually ranges from 140 to 230~K, depending on the stoichiometric amount of Fe atom\cite{doi:10.1002/ejic.200501020,doi:10.1021/acs.inorgchem.5b01260,Yi2016,
PhysRevB.93.014411,sr7_6184,PhysRevB.96.144429,Tan2018,PhysRevB.100.104403,
PhysRevB.100.134441,doi:10.1021/acs.nanolett.9b03316}. It exhibits abundant fascinating physical properties, such as large anomalous Hall effect induced by topological nodal lines\cite{PhysRevB.96.134428,kim2018large,PhysRevB.97.165415,PhysRevB.100.134441}, topological spin texture\cite{doi:10.1021/acs.nanolett.9b03453}, Kondo-lattice physics\cite{Zhangeaao6791,doi:10.1021/acs.nanolett.1c01661}, and large effective electron mass renormalization and electronic correlations\cite{PhysRevB.93.144404,PhysRevB.102.161109}. More intriguingly, the realization of a stable and tunable two-dimensional (2D) ferromagnetism even down to monolayer makes it a great platform for developing new spintronic devices based on magnetic vdW heterostructures\cite{fei2018two,Deng2018,wang2018tunneling,wang2019current,
Tan2018,PhysRevLett.125.047202,PhysRevLett.125.267205}. For insulating vdW  ferromagnets like Cr$_2$Ge$_2$Te$_6$~(Ref.~\onlinecite{Gong2017}) and CrI$_3$~(Ref.~\onlinecite{Huang2017}), the ferromagnetic state can be well described by an anisotropic Heisenberg model with local moments correlating with each other via ferromagnetic interactions. Assisted by the inherent magnetocrystalline anisotropy along the $c$ axis, thermal fluctuations can be suppressed and a 2D long-range ferromagnetic order in the monolayer limit can be stabilized\cite{Gong2017,Huang2017}. In contrast, with the involvement of electronic itinerancy, the microscopic origin of magnetism in Fe$_3$GeTe$_2$ becomes controversial.

In Ref.~\onlinecite{doi:10.7566/JPSJ.82.124711}, it was found that the results of magnetization measurements could be well presented with the modified Arrott plot, which gave a generalized Rhodes-Wohlfarth ratio one order of magnitude less than that for a local-moment system. It indicated the itinerant nature of magnetism in Fe$_3$GeTe$_2$. In Ref.~\onlinecite{Deng2018}, the observation of gate-tunable ferromagnetism was attributed to the large variation in the density of states at the Fermi level, consistent with an itinerant-electron Stoner model\cite{PhysRevB.93.134407}. On the other hand, such a model was challenged in a recent angle-resolved photoemission spectroscopy (ARPES) study in Ref.~\onlinecite{PhysRevB.101.201104}, where the electronic bands were barely changed across $T_{\rm C}$. Instead, the crucial role of local moments in Fe$_3$GeTe$_2$ was emphasized\cite{PhysRevB.101.201104}. Furthermore, an effective local-moment Hamiltonian was determined by an inelastic neutron scattering (INS) study in Ref.~\onlinecite{PhysRevB.99.094423}. In a local picture, it was also argued that the rise of the Curie temperature upon gate tuning was due to the reduction in the magnetic frustration of different exchange paths\cite{PhysRevB.103.085102}. Considering the contradictory evidence, further work is needed to clarify the issue regarding the microscopic origin of magnetism in Fe$_3$GeTe$_2$ in particular and in transition-metal compounds with $d$ electrons in general.


In this work, we use INS to study the spin dynamics in single crystals of Fe-deficient \fgt~{\new($x=0.28$)}. The excitation spectra are dominated by the high-energy ($E$) non-dispersive column-like mode, which can be attributed to the particle-hole continuum of the excitations between the exchange-split bands. There also exists a low-$E$ dispersive mode which can be attributed to collective spin excitations. We find that the spin excitations are robust against temperatures and the low-$E$ dispersive mode becomes more coherent at 100~K. Such an unusual temperature effect, which is opposite to that of a conventional Heisenberg ferromagnet, is related to the weakening of the Kondo screening effect upon warming. Our results in {\new\fgtn} can be interpreted as that both local and itinerant moments contribute to the magnetic excitations, and these two degrees of freedom couple with each other through the Kondo effect. These results demonstrate that \fgt is an excellent platform to investigate the interplay between local spins and itinerant electrons.

\section{Experimental Details and Characterizations}

Single crystals of Fe$_3$GeTe$_2$ could be grown by the chemical vapor transport method\cite{doi:10.7566/JPSJ.82.124711}. The raw powder materials Fe, Ge and Te were well mixed in a stoichiometric molar ratio of 3:1:2, with additional iodine as the transport agent. The crystals grown by this method had a negligible Fe deficiency, with a relatively higher $T_{\rm C}$~$\sim$220~K, as shown in Supplementary Fig.~1~(Ref.~\onlinecite{sm}). However, the dimension and mass of the crystals were small. Large single crystals could be obtained using the self-flux method, with the mixed Fe, Ge and Te in a molar ratio of 2:1:4, as described in Ref.~\onlinecite{PhysRevB.93.014411}. The crystals so grown, labeled as \fgt with $x$ representing about {\new$28\pm5\%$} Fe vacancies mainly on Fe(2) sites\cite{PhysRevB.99.094423,PhysRevB.93.014411}, had a reduced $T_{\rm C}$~$\sim$160~K as visualized from the susceptibility data as shown in Fig.~\ref{fig:characterizations}(d).  We label these samples as {\new\fgtn hereafter  (See Supplementary Fig.~2 for details)}. Crystals grown by both methods have an easy axis along the $c$-axis. However, the magnetic anisotropy in Fe-deficient samples is reduced compared to that in Fe$_3$GeTe$_2$. Both of these two points can be revealed readily from the difference in the magnitude of susceptibility and saturation magnetic field for field perpendicular and parallel to the $a$-$b$ plane, as shown in Fig.~\ref{fig:characterizations}(d) and Supplementary Fig.~1. Furthermore, under zero-field-cooling (ZFC) condition, the susceptibility drops significantly below a $T^*$ of $\sim$90~K for the 0.01-T field applied along the $c$-axis, which we believe to signify the antiferromagnetic correlation between the local spins and itinerant electrons, as we will explain later in this work. Despite the Fe deficiencies, no significant changes were found for the space group and lattice parameters, and the long-range ferromagnetic order\cite{PhysRevB.93.014411}. Furthermore, the relatively large mass and air-stable property made the crystals grown by the self-flux method feasible for INS experiments. Therefore, we used single crystals grown by the self-flux method for the INS measurements in this work.

\begin{figure*}[htb]
\centering
\includegraphics[width=0.9\linewidth]{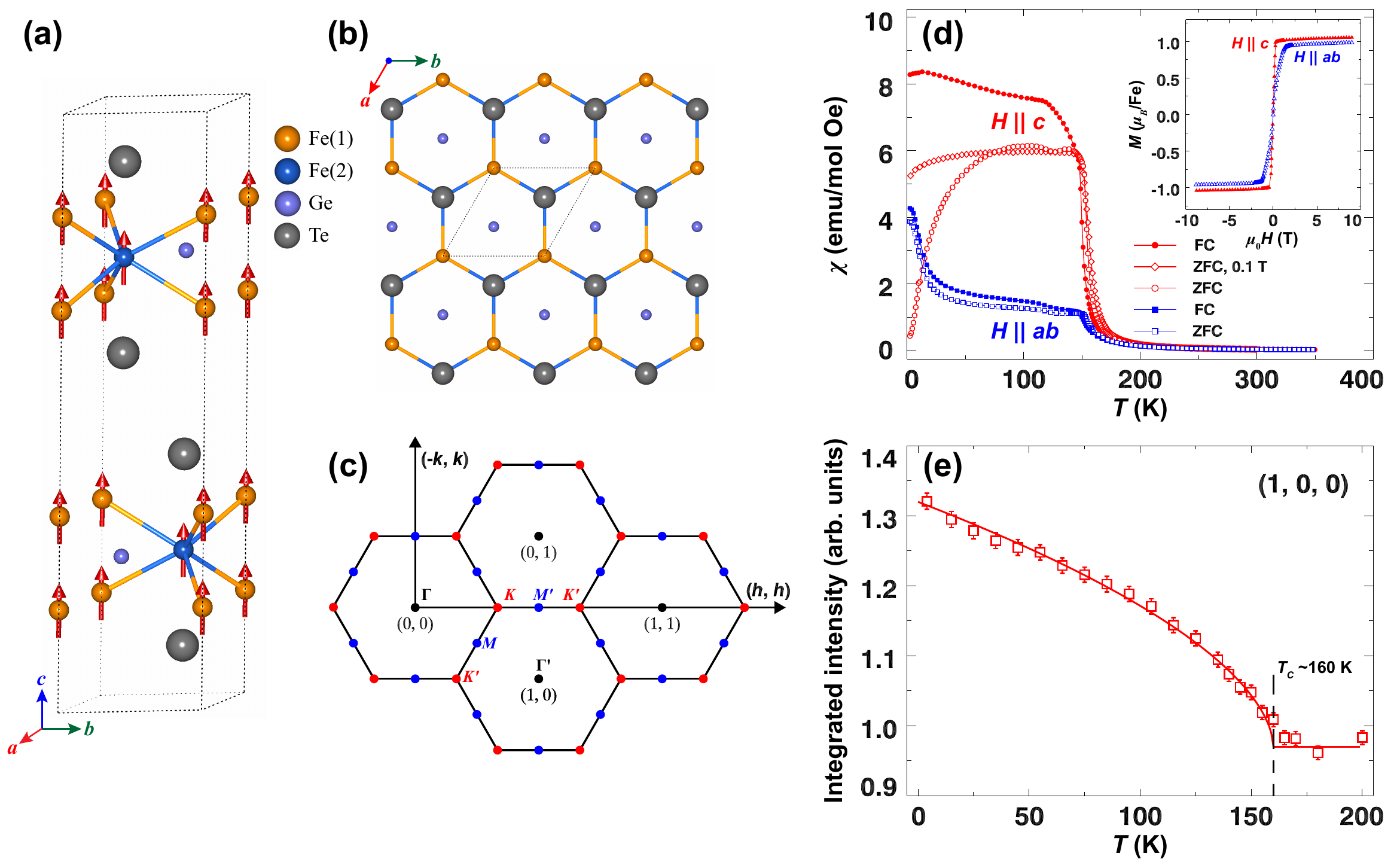}
\caption{\label{fig:characterizations}{(a) Schematic crystal and magnetic structures of Fe$_{3}$GeTe$_{2}$. There are two inequivalent sites of Fe atoms, labeled as Fe(1) and Fe(2), respectively. Arrows indicate the magnetic moments on Fe atoms. (b) Top view of the hexagonal structure in the $a$-$b$ plane. For clarity, only one Fe$_{3}$Ge slab sandwiched by two Te layers is shown. (c) Two-dimensional Brillouin zones with high-symmetry points in the ($h$,\,$k$) plane. (d) Temperature dependence of the magnetic susceptibility on a single crystal of {\new\fgtn}. Empty and filled circles or squares represent the magnetization measurements under zero-field-cooling (ZFC) and field-cooling (FC) conditions, respectively. The inset shows magnetization $M$--$H$ curves at $T=2.5$~K, with the magnetic field applied along the $c$ axis and in the $a$-$b$ plane. {\new The magnetic field applied was 0.01~T, unless marked otherwise.} (e) Temperature dependence of the integrated intensities of the magnetic Bragg peak (1,\,0,\,0) measured on SIKA, plotted in an arbitrary unit (a.u.). The error bars represent one standard deviation throughout the paper. The solid line is a guideline to the eye. The dashed line indicates the Curie temperature $T_{\rm C}$ of $\sim$160~K.}}
\end{figure*}

INS measurements were performed on 4SEASONS, a time-of-flight spectrometer at the MLF of J-PARC in Japan\cite{doi:10.1143/JPSJS.80SB.SB025}; and SIKA, a cold-neutron triple-axis spectrometer at the OPAL of ANSTO in Australia\cite{Wu_2016}. The sample array was consisted of 25 pieces of single crystals weighing about 3.95~g in total. They were coaligned and glued on aluminum plates by a backscattering Laue X-ray diffractometer. For the measurements on 4SEASONS, the sample array was mounted in a closed-cycle refrigerator in a manner that the $(h,\,h,\,l)$ plane was the horizontal plane and [-1,\,1,\,0] direction was aligned vertically. We chose a primary incident energy $E_{\rm i}=90$~meV and a chopper frequency of 350~Hz. {\new This setup gave an energy resolution of 5.12~meV [full width at half maximum (FWHM)] at the elastic line.} Since 4SEASONS was operated in a multiple-$E_{\rm i}$ mode, it had a set of other $E_{\rm i}$s of 8.9, 48.4 and 222~meV besides the primary one. {\new The respective energy resolution for these $E_{\rm i}$s are 0.35, 2.20, 17.18~meV at the elastic line.} Measurements were done at three temperatures of 4, 100 and 250~K. Scattering data were collected by rotating the sample about the [-1,\,1,\,0] direction. Raw data were reduced and combined together into four-dimensional matrices, and analyzed using the software suite Utsusemi\cite{inamura2013development}. For the measurements on SIKA, data were collected in the $(h,\,k,\,0)$ plane using a fixed final-energy mode with $E_{\rm f}=5.0$~meV. Better instrumental resolution was obtained with $E_{\rm f}=3.5$~meV, and measurements were performed in the $(h,\,h,\,l)$ plane with this setup. We used a hexagonal structure with the refined lattice parameters in Ref.~\onlinecite{PhysRevB.93.014411}, with $a=b=3.9536(7)~{\rm \AA}$ and $c=16.396(2)~{\rm \AA}$. The wave vector ${\bm Q}$ is expressed as $(h,\,k,\,l)$ in reciprocal lattice unit (rlu) of $(a^*,\,b^*,\,c^*)=(4\pi/\sqrt{3}a,\,4\pi/\sqrt{3}b,\,2\pi/c)$. To eliminate the influence of the Bose statistics, the measured neutron scattering intensities were divided by the Bose factor via $\chi''({\bm Q},\,E)=(1-{\rm e}^{-E/k_{\rm B}T})S({\bm Q},\,E)$, where $k_{\rm B}$ was the Boltzmann constant.

\section{Results}
\subsection{Ferromagnetic state and particle-hole continuum}

\begin{figure*}[htb]
\centering
\includegraphics[width=0.9\linewidth]{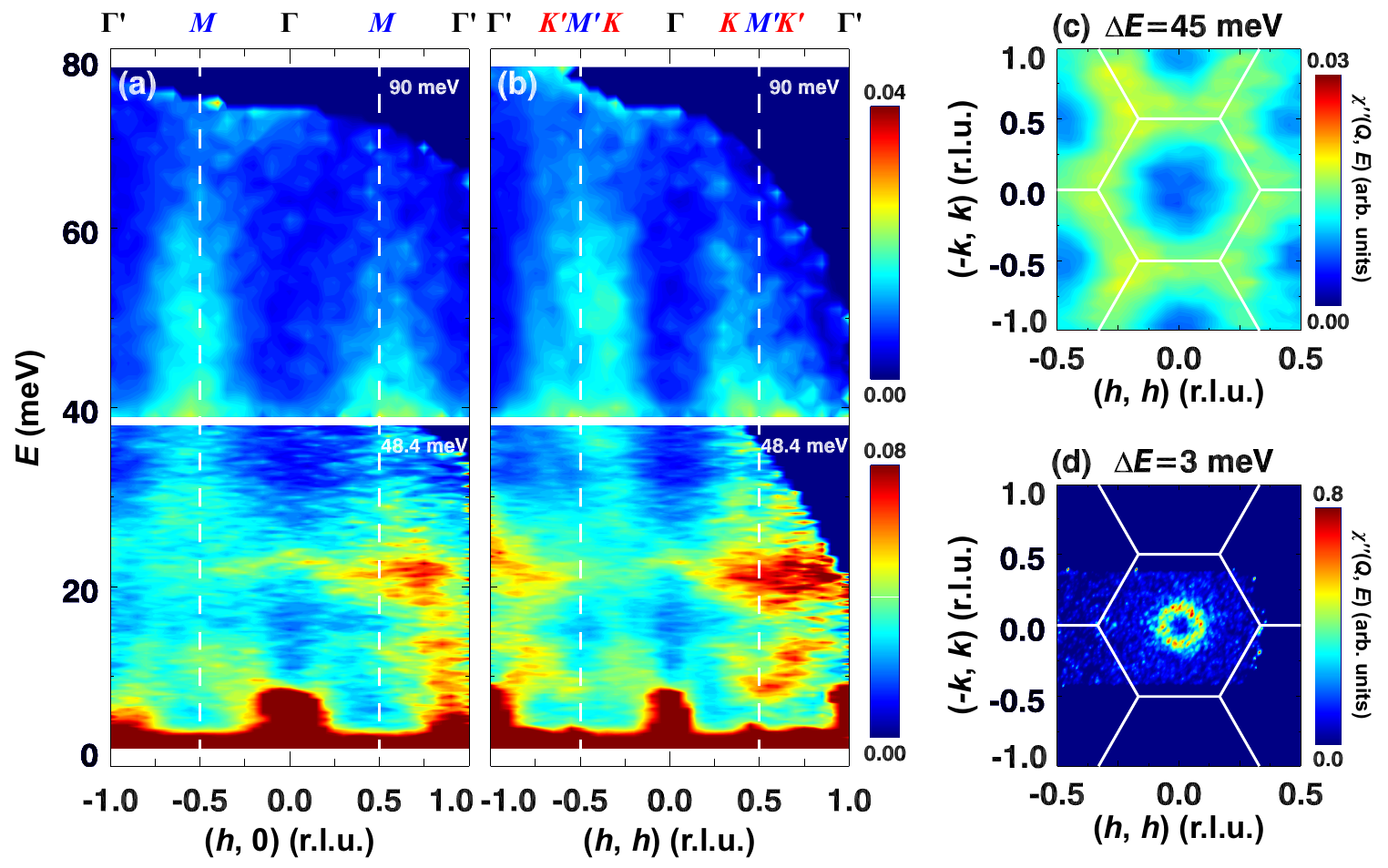}
\caption{\label{fig:fulldispersion}{(a) and (b) INS results for the spin excitation spectra at 4~K, along two high-symmetry directions [1,\,0] and [1,\,1], respectively. The lower panels in (a) and (b) were obtained with $E_{\rm i}=48.4$~meV, and integrated over $-0.1{\leq}k{\leq}0.1$~rlu and $-8{\leq}l{\leq}8$~rlu. The upper panels were obtained with $E_{\rm i}=90$~meV, and integrated over $-0.1{\leq}k{\leq}0.1$~rlu and $-10{\leq}l{\leq}10$~rlu. Vertical dashed lines indicate the M point. (c) Constant-$E$ contour at 45~meV in the ($h$,\,$k$) plane, obtained with $E_{\rm i}=90$~meV. The energy interval was chosen to be $\pm5$~meV and $l$ was integrated over $-10{\leq}l{\leq}10$~rlu. (d) Constant-$E$ contour at 3~meV in the ($h$,\,$k$) plane, obtained with $E_{\rm i}=8.9$~meV. The energy interval is $\pm$0.1~meV and $l$ was integrated over $1.8{\leq}l{\leq}2.2$~rlu. The data had been folded about the $l$ direction for better statistics. Solid lines in (c) and (d) indicate the Brillouin zone boundary.}}
\end{figure*}

The temperature dependence of the integrated intensities for the magnetic Bragg peak (1,\,0,\,0) measured on SIKA is plotted in Fig.~\ref{fig:characterizations}(e). It clearly shows that the magnetic order has an onset at the $T_{\rm C}$~$\sim$160~K for the experimental sample {\new\fgtn}. Since there is a direct relationship between $T_{\rm C}$ and Fe deficiency\cite{doi:10.1002/ejic.200501020,PhysRevB.93.014411,PhysRevB.100.104403,PhysRevB.100.134441}, the single transition indicates that our coaligned single crystals grown with the self-flux method have controllable vacancy concentration of Fe ions with {\new$x\approx0.28$}. We find no obvious temperature-dependent behavior for the intensities at the magnetic Bragg peak (0,\,0,\,2). It is expected since this wave vector is parallel to the direction of the ordered moments. It indicates that the easy axis is along the $c$ axis in \fgtn. These results are consistent with the magnetic susceptibility measurements on a piece of single crystal shown in Fig.~\ref{fig:characterizations}(d).

Figure~\ref{fig:fulldispersion}(a) and (b) show the overall profiles of the spin excitation spectra along high-symmetry directions [1,\,0] and [1,\,1] in the $(h,\,k)$ plane, respectively. These directions are defined in Fig.~\ref{fig:characterizations}(c). The upper and lower panels in Fig.~\ref{fig:fulldispersion}(a) and (b) were obtained with $E_{\rm i}=90$ and 48.4~meV, respectively, measured on 4SEASONS. There are two distinct patterns of spin excitations. First, it is found that spin-wave-like excitations disperse up from the zone center $\Gamma$ point and propagate towards the zone boundary in the lower panels of Fig.~\ref{fig:fulldispersion}(a) and (b). It indicates the ferromagnetic correlations in \fgtn. This low-$E$ dispersive mode can be better resolved using a small $E_{\rm i}=8.9$~meV, and will be discussed in detail latter. Second, we find that the dispersive mode merges into a broad continuum above $\sim$20~meV. The fluctuation continuum has a column-like shape and extends to high energies around the zone boundary, as shown in the upper panels of Fig.~\ref{fig:fulldispersion}(a) and (b). To acquire sufficient scattering intensities of the in-plane magnetic excitations, another wave vector $l$, perpendicular to the  $(h,\,k)$ plane, was integrated over large ranges in Fig.~\ref{fig:fulldispersion}(a) and (b). Such integrations can result in some contamination by phonons in the spin excitation spectra. Nevertheless, the contamination is only significant in the energy window below $\sim$25~meV, where the phonon excitations are intensive, and thus does not affect the column-like spectra at high energies (See Ref.~\onlinecite{sm} for details). We plot the constant-$E$ contours at two typical energies in Fig.~\ref{fig:fulldispersion}(c) and (d). The two characteristics of the spin excitations can be also visualized in the momentum space of the $(h,\,k)$ plane. At an energy transfer of ${\Delta}E=3\pm0.1$~meV, we find ring-like excitations around the zone center [Fig.~\ref{fig:fulldispersion}(d)], which are originated from the low-$E$ dispersive mode. As the energy increases, such excitations gradually propagate towards the Brillouin zone boundary. At ${\Delta}E=45\pm5$~meV, we find they already turn into a broad continuum spreading along the zone boundary [Fig.~\ref{fig:fulldispersion}(c)]. The broad continuum exists in a large energy window, and is still visible up to 100~meV (More contours can be found in {\new Supplementary Fig.~6}).

\begin{figure*}[htb]
\centering
\includegraphics[width=0.9\linewidth]{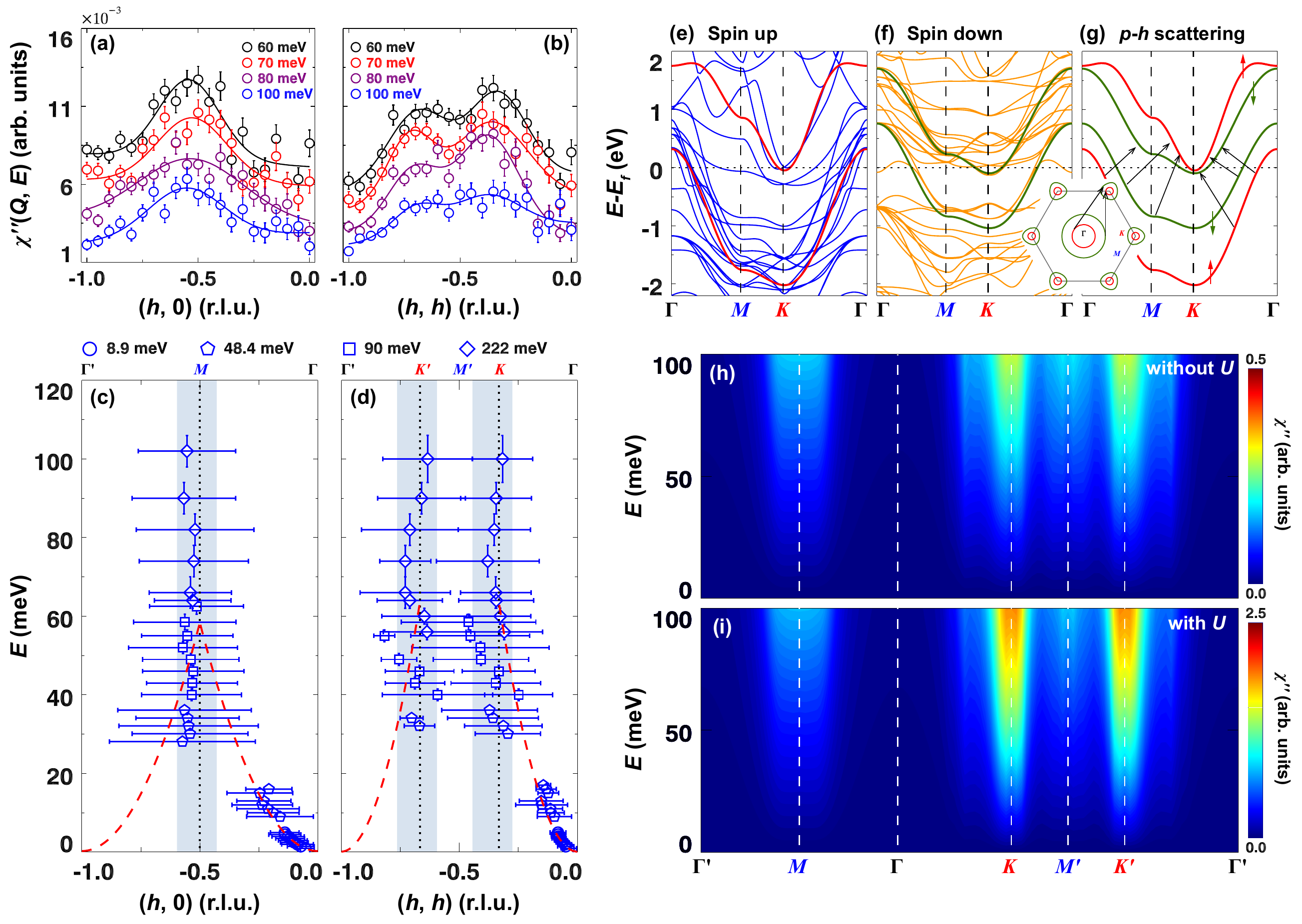}
\caption{\label{fig:spindynamics}{(a) and (b) Constant-$E$ scans at various energies along [1,\,0] and [1,\,1] directions, respectively, with $E_{\rm i}=222$~meV. For each scan, the energy intervals at selected energies were chosen to be $\pm$5~meV, and $l$ was integrated over $-10{\leq}l{\leq}10$~rlu. Solid lines are fits with Gaussian functions. Error bars represent one standard deviation throughout the paper. (c) and (d) Spin dispersions along [1,\,0] and [1,\,1] directions, respectively. The peak centers were extracted from fittings to the constant-$E$ scans as shown in (a) and (b). The horizontal error bars represent the full-width at half maximum (FWHM) of the constant-$E$ scans with Gaussian fitting, and the vertical error bars represent the energy integration interval. Dashed lines are quadratic fits to the data below 20~meV. Dotted lines indicate the Brillouin zone boundaries. Shaded regions are guides to the eye. {\new (e) and (f) DFT calculations of spin-up and spin-down bands in the ferromagnetic state, respectively. (g) Schematic diagram for the particle-hole (p-h) scattering processes. The red and green bands with up and down spins respectively are simulated bands via tight-binding model on a triangular lattice. These bands are plotted over (e) and (f) for comparisons. Black arrows represent the p-h scattering occurring between bands with different spin directions. The inset shows the corresponding Fermi surface and the wave vectors corresponding to the most significant scattering near the Fermi level. Calculated bare (h) and RPA renormalized (i) spin excitation spectra below 100~meV induced by the p-h scattering processes in (g).}}}
\end{figure*}

To quantitatively analyze the spin dynamics in {\new\fgtn}, we cut through the spin excitation spectra in Fig.~\ref{fig:fulldispersion} at various energies with appropriate $E_{\rm i}$. Examples of constant-$E$ scans through the high-$E$ broad continuum along [1,\,0] and [1,\,1] directions are shown in Fig.~\ref{fig:spindynamics}(a) and (b), respectively. Although the non-dispersive spin fluctuations exhibit continuous feature in energy, their scattering intensities actually have momentum dependence. The scans with various energies along the [1,\,0] direction show that the excitations are centered around (-0.5,\,0) in Fig.~\ref{fig:spindynamics}(a), which is the M point of the Brillouin zone. It can be found from the scans along the [1,\,1] direction in Fig.~\ref{fig:spindynamics}(b), that the scattering intensities are stronger around the K point $(-0.5\pm0.17,\,-0.5\pm0.17)$ than around the M$^\prime$ point (-0.5,\,-0.5). We extract the peak centers from the Gaussian fits to these constant-$E$ scans with various energy transfers at different $E_{\rm i}$s, and plot the peak positions in Fig.~\ref{fig:spindynamics}(c) and (d) to show the spin dispersions along the [1,\,0] and [1,\,1] directions, respectively. The low-$E$ dispersive mode is also included.

The dispersions in Fig.~\ref{fig:spindynamics}(c) and (d) showing essentially non-dispersive magnetic excitations in a large energy window are clearly beyond general spin waves. In some other magnets, where spin excitations are steep\cite{PhysRevLett.102.187206,zhao2009spin,PhysRevB.101.134408}, the excitations from the magnetic Bragg peak are seemingly dispersionless in the low-$E$ window, but will disperse outward and become broad at high energies\cite{PhysRevLett.102.187206,zhao2009spin,PhysRevB.101.134408}. {\new This is certainly not the case here, as the low-$E$ excitations are dispersive while the high-$E$ excitations are column-like. A quadratic fitting, which can describe the ferromagnetic spin waves to some extent, is applied to the low-$E$ dispersive mode along the [1,\,0] direction in Fig.~\ref{fig:spindynamics}(c). It reaches its maximum energy of $\sim$60~meV at the M point, but cannot contribute to the magnetic scattering above it. For the [1,\,1] direction, although the tendency of the fit can reach a higher energy at the ${\rm M}'$ point, the real spin-wave energy should be the same for the symmetry-equivalent ${\rm M}'$ and M points. Furthermore, the column-like excitations around the Brillouin zone boundary actually can stem from very low energy. This can be more clearly visualized from the excitations along the $c$-axis shown in Supplementary Fig.~5. It is worth mentioning that there exists a significant amount of $\sim$28\% Fe deficiency in our sample, which will introduce the disorder effect into the system\cite{PhysRevB.99.094423}. Additionally, magnon-magnon interactions can also damp the spin waves and make them broader. However, the disorder effect will not change the profile of the spin waves, and the magnon-magnon interactions usually occur at energies above the single-magnon excitations. The column-like excitations in the entire energy range at the Brillouin zone boundary are inconsistent with the ferromagnetic spin waves dispersing from the Brillouin zone center, even after considering the disorder effect and magnon-magnon interactions. The conflict between our results and the spin-wave interpretation in a local-moment Heisenberg model urges us to consider the role of itinerant electrons\cite{doi:10.7566/JPSJ.82.124711,Deng2018,Zhangeaao6791}.}

Within the framework of itinerant Stoner model, ferromagnetic state is originated from the unbalanced numbers of spin-up and spin-down electrons induced by the exchange splitting\cite{S_1947,Feng2013,PhysRevB.99.014407}. {\new In such a case, spin-flip particle-hole excitations can happen between the spin-up and the spin-down electronic bands, as shown schematically in Fig.~\ref{fig:spindynamics}(g). For these spin excitations, with the same transferred wave vector $\bm Q$, multiple particle-hole excitations with various transferred energies can be permitted. That can result in a particle-hole continuum in the itinerant ferromagnetic excitations\cite{S_1947,Feng2013,PhysRevB.99.014407}. The profile of this continuum is determined by the exchange splitting energy $\Delta$ and the specific band structures. To figure out the origin of the column-like excitations, we firstly use the density-functional theory (DFT) to calculate the electronic bands of this material\cite{Giannozzi_2017}. The results in Fig.~\ref{fig:spindynamics}(e) and (f) show that in the ferromagnetic state, the spin-up and spin-down bands are separated by a large exchange splitting $\Delta\sim$1.5~eV, consistent with that in Ref.~\onlinecite{PhysRevB.101.201104}. From the band structures, we find that the dispersions of the bands resemble those of a triangular lattice. To capture the major characteristics of the bands across the Fermi level, we next apply a tight-binding model on a triangular lattice consisting of four electronic bands as shown in Fig.~\ref{fig:spindynamics}(g) to simulate the multiband nature in the material. It can be found that the simulated bands (red for spin up and green for spin down) are similar to the DFT results shown in Fig.~\ref{fig:spindynamics}(e) and (f). Moreover, the Fermi surface formed by these bands as shown in the inset of Fig.~\ref{fig:spindynamics}(g), is consistent with the observed Fermi surface by the ARPES measurements\cite{Zhangeaao6791,PhysRevB.101.201104}, where there are two hole pockets near the $\Gamma$ point and a small electron pocket near the K point. Finally, we use the random phase approximation (RPA) to calculate the magnetic excitation spectra resulting from the particle-hole scattering processes between these four bands as sketched in Fig.~\ref{fig:spindynamics}(g). The wave vectors corresponding to the strongest scatterings are illustrated in the inset of Fig.~\ref{fig:spindynamics}(g). The calculated magnetic excitation spectra resulting from these scattering processes without and with the correction of the Hubbard interaction $U$ are shown in Fig.~\ref{fig:spindynamics}(h) and (i), respectively. It can be found that the main features in the experiment that the column-like excitations occur at the zone boundary and the intensities are stronger at the K point than at the M point as shown in Fig.~\ref{fig:fulldispersion}(a) and (b), are reproduced. Moreover, since there is no qualitative difference between the profiles of the spin excitation spectra with and without $U$, our analysis does not depend on the specific value of $U$. {\newr We note that the intensities are higher at high energies than those at low energies, different from the experiment. We think this is because, in addition to the spin-flip particle-hole excitations, the low-energy intensities in the experimental spectra also have contributions from spin waves (as we discuss below) as well as phonons, while the RPA calculations only consider the spin excitations from the itinerant electrons. Besides, this is a multiband system, while in the calculations, we only use simplified four bands as shown in Fig.~\ref{fig:spindynamics}(g) to capture the main features. By considering contributions from all the bands crossing the Fermi level together with spin waves and phonons, the level of agreement on the intensity between the experiment and calculations may be improved.} With these results, we can attribute the column-like excitations to the particle-hole excitation continuum between the exchange-split bands (More details about the DFT and RPA calculations can be found in Ref.~\onlinecite{sm}). Such a continuum has been hypothesized theoretically for a long time\cite{S_1947,Feng2013,PhysRevB.99.014407}, but rarely observed experimentally\cite{npjqm6_60,nc11_3076}. It reflects the itinerant ferromagnetism induced by the exchange splitting in \fgtn.}

\subsection{Low-$E$ spin waves and Kondo effect}

\begin{figure*}[htb]
\centering
\includegraphics[width=0.8\linewidth]{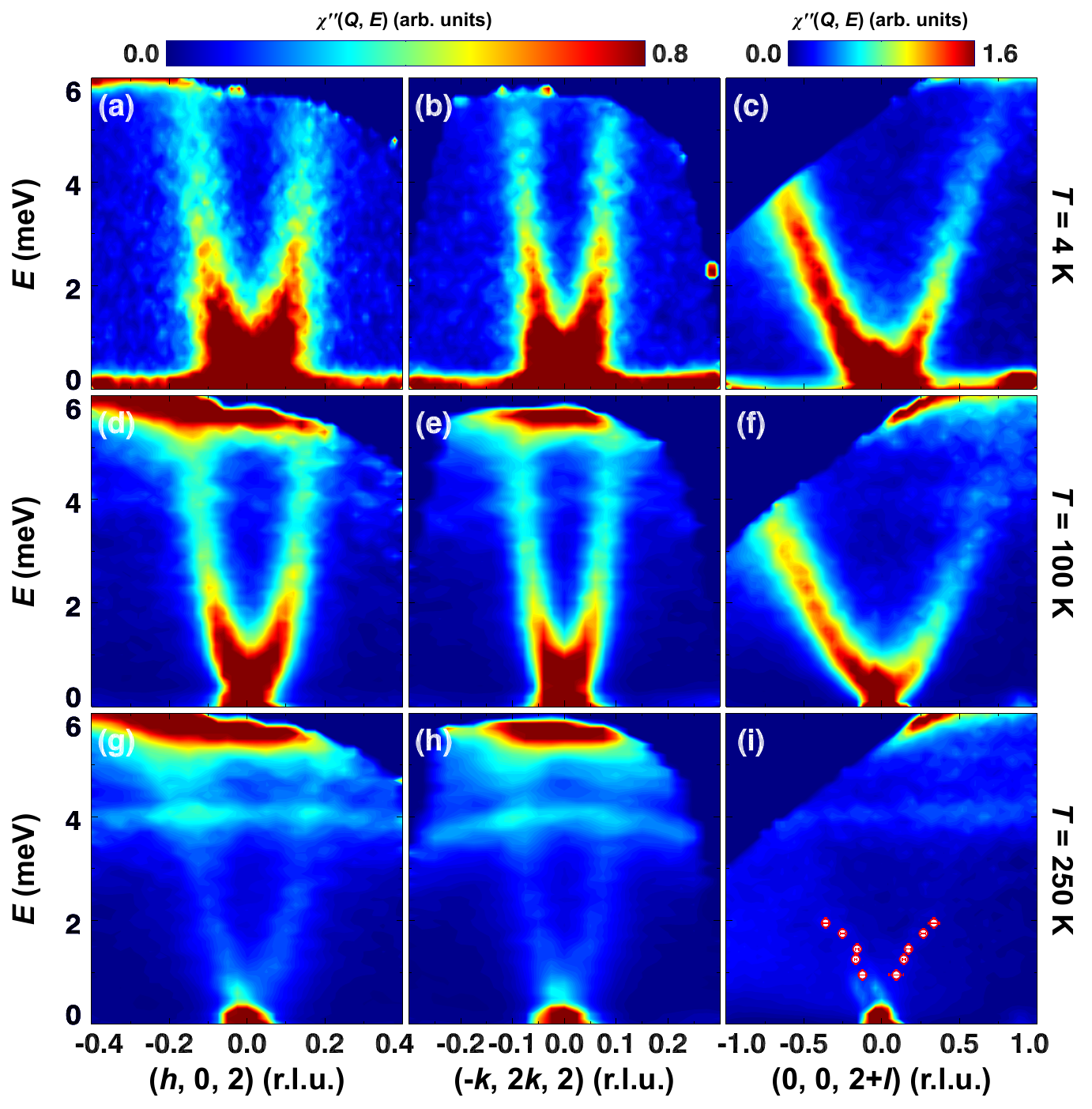}
\caption{\label{fig:lowedispersion}{Low-$E$ spin excitation spectra obtained with $E_{\rm i}=8.9$~meV at 4~K [(a)-(c)], 100~K [(d)-(f)], and 250~K [(g)-(i)], respectively. The spectra in the left [(a), (d) and (g)], middle [(b), (e) and (h)] and right [(c), (f) and (i)] columns represent the spin excitations along [1,\,0,\,0], [-1,\,2,\,0] and [0,\,0,\,1] directions, respectively. The integration ranges of $k$, $h$ and $l$ when necessary were $-0.04{\leq}k{\leq}0.04$~rlu, $-0.05{\leq}h{\leq}0.05$~rlu and $1.8{\leq}l{\leq}2.2$~rlu. The data have been folded about the $l$ direction to improve the statistics. Data points in (i) were obtained by fitting the constant-$E$ cuts shown in Supplementary Fig.~9.}}
\end{figure*}

\begin{figure*}[htb]
\centering
\includegraphics[width=0.85\linewidth]{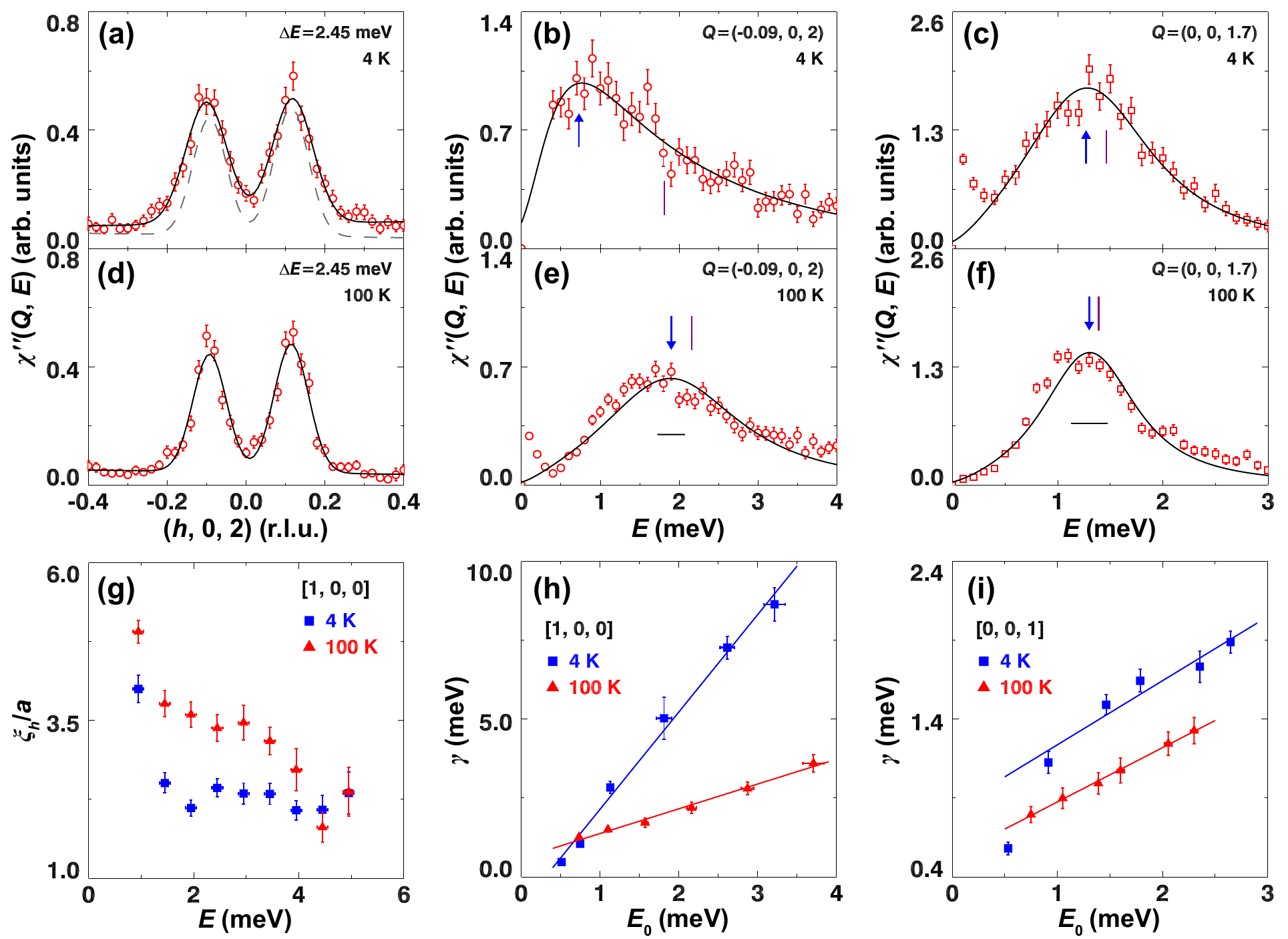}
\caption{\label{fig:tdependence}{{\new (a) Constant-$E$ scans at 2.45~meV along the [1,\,0,\,0] direction, measured at 4~K and (d) at 100~K {\newr with $E_{\rm i}=8.9$~meV and a resulting energy resolution of 0.35~meV.} The energy interval for the integration is $\pm 0.1$~meV. Solid curves are fits with Gaussian functions convoluted with the instrumental resolutions. The dashed curve in (a) is the fitting result of (d), and is plotted for comparison purpose. (b) Constant-${\bm Q}$ cuts at (-0.09,\,0,\,2), measured at 4~K and (e) at 100~K. The thickness of $h$ is 0.02~rlu. (c) Constant-${\bm Q}$ cuts at (0,\,0,\,1.7), measured at 4~K and (f) at 100~K. The thickness of $l$ is 0.04~rlu. Solid curves in these Constant-${\bm Q}$ cuts are fits with the DHO formula convoluted with the instrumental resolutions as discussed in the text. Arrows mark the peak center of the fitted DHO, and the vertical bars mark the magnon energy $E_0$. Horizontal bars in (e) and (f) represent the instrumental resolutions at the elastic line. (g) Dynamic spin-spin correlation lengths extracted from fitting to the constant-$E$ scans and normalized to the lattice parameter $a$ for the [1,\,0,\,0] direction. (h) and (i) Damping constant $\gamma$ as a function of magnon energy $E_0$ at two temperatures for [1,\,0,\,0] and [0,\,0,\,1] directions, respectively. The horizontal bars in (g) represent the energy integration interval of $\pm$0.1~meV, and in (h) and (i) represent the fitting errors for $E_0$. Solid lines in (h) and (i) are guides to the eye.}}}
\end{figure*}

{\new To better characterize the low-$E$ dispersive mode in {\new\fgtn}, we use the high-resolution data with $E_{\rm i}=8.9$~meV obtained on 4SEASONS.} In Fig.~\ref{fig:lowedispersion}, we plot the excitation spectra at 4~K for both in-plane [Fig.~\ref{fig:lowedispersion}(a) and (b)] and out-of-plane [Fig.~\ref{fig:lowedispersion}(c)] directions. We chose the [-1,\,2,\,0] direction symmetrically equivalent to the [1,\,1,\,0] direction used in Figs.~\ref{fig:fulldispersion} and \ref{fig:spindynamics} to avoid spurious peaks at low energies. Besides the steep in-plane dispersions, a prominent out-of-plane dispersion is also observed. From the spectra in Fig.~\ref{fig:lowedispersion}, no spin gap is observed. As shown in Supplementary Fig.~8(a), the constant-$\bm Q$ cut at $(0,\,0,\,2)$ does not show any features corresponding to the spin gap either, even in the measurements with much finer instrumental resolution performed on SIKA. Instead, additional quasielastic scattering is found. This result is contrary to a spin gap of 3.7~meV reported by a previous INS experiment in Ref.~\onlinecite{PhysRevB.99.094423}. Actually, the ring-like excitations at $3\pm0.1$~meV in Fig.~\ref{fig:fulldispersion}(d) also evidence an overestimated value of 3.7~meV about the spin gap. We do not think the samples of these two experiments have an essential difference, because of the same growth method, similar Fe deficiency, and close $T_{\rm C}$. The reason for the different results can be that the out-of-plane dispersion was ignored in Ref.~\onlinecite{PhysRevB.99.094423}, where the in-plane dispersion and the spin gap were determined with a too-wide integration range of $l$ over $-2{\leq}l{\leq}2$~rlu. In fact, with our data, we can also obtain exactly the same spin ``gap" value of 3.7~meV using the same integration range, as demonstrated in Supplementary Fig.~7.

The high-resolution data allow us to examine the low-$E$ spin-wave-like excitations quantitatively. In the small wave-vector limit, the ferromagnetic spin waves can be described by a well-known quadratic dispersion relation $E=E_{\rm g}+D{\bm Q}^2$, where $E_{\rm g}$ is the spin gap and $D$ is the spin-wave stiffness. It gives $D$ of $69.0\pm1.8$~meV~\AA$^2$ for the [1,\,0,\,0] direction, $56.7\pm2.9$~meV~\AA$^2$ for the [-1,\,2,\,0] direction and $53.6\pm1.9$~meV~\AA$^2$ for the [0,\,0,\,1] direction (See Supplementary Fig.~8 for details). The comparable values of the spin-wave stiffness for in-plane and out-of-plane directions suggest the three-dimensional nature of the magnetism in {\new\fgtn}, despite its quasi-two-dimensional vdW lattice structure. A recent ARPES study reported strong variation of the electronic bands along the out-of-plane direction\cite{PhysRevB.101.201104}. It can provide the interlayer hopping channel for the magnetic exchange interaction, whose strength can be tuned by protonic gating, as observed from the exchange-bias measurements\cite{PhysRevLett.125.047202}. The scaling analysis on \fgt also revealed a behavior of three-dimensional magnetism\cite{PhysRevB.96.144429,Tan2018}, consistent with our results. Besides, nonzero intercepts can be found in Supplementary Fig.~8(c), implying a spin gap around 0.5~meV, even though there is no evidence of a spin gap in Fig.~\ref{fig:lowedispersion} and Supplementary Fig.~8(a).

The spin gap is an manifestation of the out-of-plane magnetic anisotropy, which is believed to be crucial in stabilizing the 2D long-range ferromagetic order down to monolayer in Fe$_3$GeTe$_2$~(Refs.~\onlinecite{fei2018two,Deng2018}). It was estimated that in the monolayer limit, a single-ion anisotropy of $\sim$2~meV was required\cite{Deng2018}. On these results, we would like to make a few remarks. First, our experimental sample possesses certain Fe deficiencies that will reduce the magnetic anisotropy\cite{doi:10.1021/acs.nanolett.9b03316}, and thus reduce the spin gap as a result. It can be revealed from the difference in the magnetization saturation fields for field perpendicular to and within the $a$-$b$ plane, which is $\sim$1.5~T in our sample in the inset of Fig.~\ref{fig:characterizations}(d), compared to $\sim$5~T in a crystal with negligible deficiency in Supplementary Fig.~1. Second, the out-of-plane magnetic anisotropy can be significantly enhanced with reduced sample thickness\cite{Tan2018}. Finally, from Supplementary Fig.~8(a), we believe that there exist diffusive fluctuations induced by itinerant electrons. On the one hand, they can contribute to the quasielastic scattering and fill up the spin gap\cite{PhysRev.182.624,PhysRevLett.51.300,RevModPhys.60.209,PhysRevLett.76.4046}. On the other hand, given that the magnetic excitations are diffusive, the energy of spin waves deduced from constant-$E$ scans in Supplementary Fig. 8(c) can be overestimated\cite{PhysRevB.36.881}. This can explain the discrepancy on the absence [Fig.~\ref{fig:lowedispersion} and Supplementary Fig.~8(a)] or presence [Supplementary Fig.~8(c)] of a spin gap.

The low-$E$ dispersive mode at higher temperatures 100 and 250~K ($\sim$1.6$T_{\rm C}$) is also investigated, and plotted in Fig.~\ref{fig:lowedispersion}(d)-(f) and (g)-(i), respectively. It is found that the excitations still maintain the spin-wave-like profile upon warming. The paramagnetic scattering intensities become much weaker at 250~K, especially for those along the [0,\,0,\,1] direction. The temperature evolution of the spin-wave stiffness is plotted in Supplementary Fig.~8(d). We find that the in-plane spin excitations gradually soften upon warming, but they still persist even at 1.6$T_{\rm C}$. The out-of-plane spin excitations also soften upon warming in the ferromagnetic phase. For this direction at 1.6$T_{\rm C}$, the paramagnon dispersion can still be identified by doing various constant-$E$ scans, as shown in Supplementary Fig.~9, although the intensities are much weaker compared to those in the $a$-$b$ plane. We plot the extracted peak centers in Fig.~\ref{fig:lowedispersion}(i). Based on these data points, the spin-wave stiffness is finite. As we know, for a conventional Heisenberg ferromagnet with second order ferromagnetic phase transition, the spin-wave stiffness should vanish at $T_{\rm C}$ via $D(T)\propto(1-T/T_{\rm C})^{\nu-\beta}$, where $\nu$ and $\beta$ are critical exponents of the magnetic phase transition for the correlation length and magnetization, respectively\cite{RevModPhys.39.395,PhysRev.177.952,collins1989magnetic,PhysRevB.101.134418}. Therefore, our results showing the persistent low-$E$ dispersive mode above $T_{\rm C}$ suggest that the ferromagnetic phase transition in {\new\fgtn} is weakly first order instead of second order\cite{PhysRevB.96.144429}. In addition, the high-$E$ non-dispersive mode also changes little across $T_{\rm C}$, as shown in  Supplementary Fig.~10. {\new We notice that in some quasi-2D magnets, persistent in-plane spin waves can be observed above magnetic phase transition temperature, even though the long-range order is destroyed by thermal fluctuations\cite{PhysRevB.76.212402,PhysRevB.84.054544,PhysRevB.72.014439,PhysRevB.79.054526}. It is usually attributed to the short-range order induced by much stronger intralayer magnetic interactions\cite{PhysRevB.76.212402,PhysRevB.84.054544,PhysRevB.72.014439,
PhysRevB.79.054526}. However, such a picture in the low-dimensional system may not be applicable to understand the temperature evolution of \fgtn. First, despite the quasi-2D vdW structure, \fgtn actually has a 3D nature of magnetism, because of the comparable spin-wave stiffness for in-plane and out-of-plane directions as discussed above. Second, both in-plane and out-of-plane spin excitations persist above $T_{\rm C}$ as shown in Fig.~\ref{fig:lowedispersion}(g)-(i).}

{\new By bare-eye inspections, it can be found that the excitations at 100~K [middle panels of Fig.~\ref{fig:lowedispersion}] are sharper both along the energy and momentum axes than those at 4~K [top panels of Fig.~\ref{fig:lowedispersion}], especially for the in-plane directions [Fig.~\ref{fig:lowedispersion}(d)\&(e) $vs$ (a)\&(b)]. To analyze the data quantitatively, we plot constant-$E$ scans at 2.45~meV along [1,\,0,\,0] direction at 4 and 100~K in Fig.~\ref{fig:tdependence}(a) and (d), respectively. It is found that the peaks are sharper at 100~K than at 4~K. The linewidths of the peaks are related to the inverse of the dynamic spin-spin correlation lengths $\xi$. We extract $\xi$ at various energies and plot them in Fig.~\ref{fig:tdependence}(g). Overall, we find that the dynamic spin-spin correlation lengths of the low-$E$ spin waves are very short, which are only about 1.5 to 3.5 lattice constants in the $a$-$b$ plane even at low temperatures, where well-defined and long-range correlated spin waves are expected. It reflects the damping character of this mode. More intriguingly, as shown in Fig.~\ref{fig:tdependence}(g), it is clear that the low-$E$ spin waves possess longer correlation lengths at 100~K than at 4~K.}

{\new To further quantify the temperature effect, we plot the constant-${\bm Q}$ cuts through the in-plane [Fig.~\ref{fig:tdependence}(b) and (e)] and out-of-plane dispersions [Fig.~\ref{fig:tdependence}(c) and (f)] at 4 and 100~K. At ${\bm Q}=(-0.09,\,0,\,2)$, a peak of the spin excitations can only be found at 100~K, as shown in Fig.~\ref{fig:tdependence}(e). At 4~K, no such peak is observed, but instead, there is a broad hump just above zero energy in Fig.~\ref{fig:tdependence}(b). At ${\bm Q}=(0,\,0,\,1.7)$, inelastic peaks can be found at both 4 and 100~K, as shown in Fig.~\ref{fig:tdependence}(c) and (f). The peak centers do not change obviously, but the peak becomes sharper at 100~K than at 4~K. To quantitatively understand the results, we convolute the instrumental resolution with a damped harmonic oscillator (DHO) formula to fit the data, which is often used to describe damped spin waves\cite{zhao2009spin,nc11_3076}. It has a form of $\chi''({\bm Q},E)\propto\frac{{\gamma}E_0E}{(E^2-E_0^2)^2+({\gamma}E)^2}$, where $E_0$ is the magnon energy and $\gamma$ is the damping constant (the inverse of $\gamma/2$ represents the lifetime of the damped magnons). For the cuts at ${\bm Q}=(0,\,0,\,1.7)$, with the DHO fitting, we find that the actual peak center slightly shifts away from the magnon energy $E_0$ to lower energy, as shown in Fig.~\ref{fig:tdependence}(c) and (f). It is owing to the comparable values of $E_0$ and the damping constant $\gamma$. Nevertheless, a larger $\gamma$ is extracted at 4~K. For the cuts at ${\bm Q}=(-0.09,\,0,\,2)$, a similar case about the peak and $E_0$ occurs at 100~K in Fig.~\ref{fig:tdependence}(e). Moreover, the broad hump at 4~K is reproduced by the DHO fitting, with peak center much smaller than $E_0$ as shown in Fig.~\ref{fig:tdependence}(b). In this case, the damping constant $\gamma$ is much larger than the magnon energy $E_0$, so that there is no inelastic peak by the spin excitations centered around $E_0$. It represents a state of overdamped magnons (See Supplementary Note 6 for details). More constant-${\bm Q}$ cuts at different ${\bm Q}$s along [1,\,0,\,0] and [0,\,0,\,1] directions are plotted in Supplementary Figs.~11 and 12, respectively, where similar phenomena can be found. These indicate the lifetime and coherence of the dispersive mode are enhanced at 100~K. To better present this conclusion, we extract the fitted parameters $\gamma$ and $E_0$ from Supplementary Figs.~11 and 12, and plot $\gamma$ as a function of $E_0$ for [1,\,0,\,0] and [0,\,0,\,1] directions in Fig.~\ref{fig:tdependence}(h) and (i), respectively. It can be found that the damping constant $\gamma$ at 4~K is larger than that at 100~K along both directions, especially for the in-plane cuts.}

{\new Our results that the low-$E$ magnons at 100~K are sharper in both the momentum and energy than those at 4~K are totally different from the temperature evolution of normal spin excitations. The latter are well-defined and long-lived at low temperatures, and will gradually lose correlation and coherence upon warming due to the enhancement of  thermal fluctuations\cite{PhysRevLett.109.127206,Bayrakci1926,Huberman_2008}. What is the origin of the stronger damping on the low-$E$ spin excitations in \fgtn at low temperatures? One possibility is the effect of disorder as a result of the Fe deficiency. However, it is weakly energy and wave-vector dependent, and particularly is not expected to exhibit a temperature dependence, which conflicts with the results in Fig.~\ref{fig:tdependence}. For spin excitations resulting from the itinerant electrons, such an unusual temperature effect is not anticipated either. Instead, we believe that the damping and the temperature effect in \fgtn can be understood when considering the Kondo screening effect of the itinerant electrons on the local moments. Actually, the manifestation of coupled local moments and itinerant electrons has been implied by the linear relationship of $\gamma$ against $E_0$ in Fig.~\ref{fig:tdependence}(h) and (i) (Ref.~\onlinecite{nc11_3076}). Moreover, such coupling seems to be much stronger in the $a$-$b$ plane, which can result in a much heavier damping of the in-plane spin waves than those along the out-of-plane direction, as revealed by a significant increase of the slope at 4~K along [1,\,0,\,0] direction in Fig.~\ref{fig:tdependence}(h). In fact, some previous studies also reported evidence for the Kondo-lattice behavior in \fgt (Refs.~\onlinecite{Zhangeaao6791,doi:10.1021/acs.nanolett.1c01661})}. In a Kondo lattice, each local moment is coupled with and screened by the surrounding itinerant electrons. The interplay between them directly contributes to the damping character of the collective excitations of local moments. {\new That also explains the short dynamic spin-spin correlation lengths of the dispersive mode as shown in Fig.~\ref{fig:tdependence}(g)}. Moreover, the Kondo screening effect by surrounding itinerant electrons is expected to be more significant at low temperatures, and will weaken upon warming. In this material, there exists coherent-incoherent crossover in the ferromagnetic phase with a characteristic temperature $T^*$, under the interpretation of Kondo-lattice physics\cite{Zhangeaao6791,PhysRevB.102.161109}. Below $T^*$, heavy electrons are expected to emerge with further enhanced effective electron mass. Similar to the analysis in Ref.~\onlinecite{Zhangeaao6791}, $T^*$~$\sim90\pm10$~K can be obtained from the resistivity measurements in our sample of {\new\fgtn}, as shown in Supplementary Fig.~13. We think at a temperature below $T^*$, stronger coherent scattering between collective excitations and surrounding itinerant electrons occurs, which will seriously shorten the lifetime of collective excitations. Such scattering process will be reduced above $T^*$. This can explain why the dispersive mode is better defined at 100~K than at 4~K as shown in Figs.~\ref{fig:lowedispersion} and \ref{fig:tdependence}.

\section{Discussions}

Our INS results on the spin dynamics of {\new\fgtn} shown in Figs.~\ref{fig:fulldispersion} and \ref{fig:lowedispersion} clearly reveal two components of the excitation spectra, with the low-$E$ collective excitations dispersing from the $\Gamma$ point and the column-like continuum extending up to above 100~meV located at the Brillouin zone boundary. This two-component structure is observed for both in- and out-of-plane directions (Figs.~\ref{fig:fulldispersion}-\ref{fig:lowedispersion} and Supplementary Fig.~5). {\new We attribute the column-like excitations to the particle-hole continuum because: First, they conflict with the spin waves in local-moment framework. Second, they can be reproduced by the itinerant-electron model as shown in Fig.~\ref{fig:spindynamics}(h) and (i). This pattern of excitations reflects the itinerant ferromagnetism induced by the exchange splitting in \fgtn. In contrast,} in most previous studies of $d$-electron systems, the spin dynamics can be described under the framework of spin waves, regardless of itinerant electrons\cite{doi:10.1143/JPSJ.75.111002,PhysRev.179.417,PhysRevB.11.2624,PhysRev.182.624,PhysRevLett.30.556,PhysRevB.23.198,PhysRevLett.102.187206,zhao2009spin}. The experimental evidence of the abrupt decrease in magnetic scattering intensities and the spectra broadening at high energies is usually interpreted as the entering of spin waves into the particle-hole continuum\cite{PhysRevB.11.2624,PhysRevLett.30.556,PhysRevB.23.198,PhysRevLett.102.187206}. However, the presence of the continuum is mostly based on conjecture and no clear signature of the continuum has been really observed. Therefore, whether such a picture is correct is still debatable\cite{PhysRevLett.102.187206,zhao2009spin}. On the other hand, in our results, based on the two-component character of the spectra, the collective excitations and particle-hole continuum are unambiguously distinguished, providing a rare example demonstrating how the low-$E$ dispersive mode enters the continuum\cite{npjqm6_60,nc11_3076}.

{\new Although the itinerant-electron model can explain the observed column-like continuum at high energies, and seemingly the collective excitations of itinerant electrons within this model can also induce the dispersive mode at low energies\cite{S_1947,Feng2013,PhysRevB.99.014407}, it fails to produce the damping character and the unusual temperature effect in Fig.~\ref{fig:tdependence}. Instead, the Kondo-lattice model is considered. In this model, we think local moments should firstly form and align in parallel to each other through ferromagnetic interactions, which can directly contribute to the low-$E$ spin waves in Fig.~\ref{fig:lowedispersion}. In addition, the surrounding itinerant electrons are coupled with and aligned antiparallel to the local moment at each site via antiferromagnetic couplings, giving rise to the Kondo screening effect. The interplay between local moments and itinerant electrons will cause the damping of the spin waves, and give rise to the unusual behavior against temperatures in the sense that the Kondo-screening effect is enhanced at low temperatures and thus the correlation length and lifetime of the low-$E$ magnons decrease in Fig.~\ref{fig:tdependence}(g)-(i).} This is consistent with the observation of a $T^*$~$\sim90\pm10$~K obtained from the resistivity measurements in our sample of {\new\fgtn}, as shown in Supplementary Fig.~13, where coherent heavy electrons emerge below $T^*$. This picture is also strongly supported by the observation of a significant reduction of the magnetic susceptibility below $T^*$, as shown in Fig.~\ref{fig:characterizations}(d) and Supplementary Fig.~1 (ZFC condition). It also reveals that the surrounding itinerant electrons will be detrimental to the establishment of ferromagnetic ground state by the local moments. This view is strongly supported by the experimental observation that the decrease of electrical conductance by an ionic gate will dramatically elevate the $T_{\rm C}$ in Ref.~\onlinecite{Deng2018}. On the one hand, the spins of itinerant electrons are coupled with the local moment at each site and prefer to align parallel to each other. On the other hand, they also prefer to be antiparallel between nearest-neighbor sites due to the Pauli exclusion. As a result, there is a competition between the ferromagnetic and antiferromagnetic correlations for the itinerant electrons\cite{Yi2016}. From this point of view, the observed topological spin texture can be understood as resulting from the competing magnetic interactions\cite{doi:10.1021/acs.nanolett.9b03453}. It is worth mentioning that the spin frustration for the itinerant electrons can be relieved in the vicinity of a Kondo hole. This in turn will enhance the Kondo screening effect on the local moments adjacent to the Kondo hole\cite{doi:10.1021/acs.nanolett.1c01661}, and reduce the ferromagnetic interactions between the local spins. We believe that the $T_{\rm C}$ is primarily controlled by the exchange coupling between the local spins\cite{PhysRevB.99.094423,PhysRevB.103.085102}, considering the huge splitting energy $\Delta$ of $\sim$1.5~eV which is about two orders of magnitude higher than the $T_{\rm C}$. {\new Previous ARPES measurements also found the electronic bands were barely changed across $T_{\rm C}$, while the exchange splitting was expected to disappear above $T_{\rm C}$ in an itinerant-electron model\cite{PhysRevB.101.201104}.} Nevertheless, the $T_{\rm C}$ is adjustable by an ionic gate that manipulates the charge carriers through the interplay between local moments and itinerant electrons.

{\new According to these results, we consider that the combination of itinerant and local-moment pictures is essential to explain the nature of magnetism in this material\cite{Zhangeaao6791,doi:10.1021/acs.nanolett.1c01661,PhysRevB.102.161109}.} The final question is what is the microscopic origin of the local and itinerant moments in \fgtn? One possible answer is that, both of these two components are from the 3$d$ orbitals of Fe atoms. Some of them with less itinerancy will be localized due to the strong onsite Coulomb repulsion, forming local moments. {\new This conjecture is indeed supported by our orbital-resolved band calculations in Supplementary Fig.~14. We notice that in the non-magnetic state, the $d_{z^2}$ orbital has a narrower energy distribution than other orbitals and locates near the Fermi level, which will more easily lead to the opening of a Mott-Hubbard gap and correspondingly the formation of the local moments under some electron interactions. Especially, in the magnetic state, our claculations reveal that the spin-up $d_{z^2}$ orbital of the Fe(2) atom is below the Fermi level, while the spin-down $d_{z^2}$ orbital is above the Fermi level [Supplementary Fig.~14(e) and (f)], leading to an almost completely polarized $d_{z^2}$ orbital. This implies that this orbital plays the most important role in the formation of local moments. In addition, it can be found that the strongly dispersive bands near the Fermi level also come from Fe 3$d$ orbitals, which contribute to the itinerant moments in the magnetic state. Furthermore, the orbital-resolved band structures also reveal that the localized orbitals have distinct hybridizations with some strongly dispersive bands near the Fermi level, which can induce the formation of the Kondo singlets between the local moments and itinerant electrons\cite{Zhangeaao6791,PhysRevB.103.L060403}. We consider our results reflect the dual properties of the 3$d$ electrons\cite{Zhangeaao6791,doi:10.1021/acs.nanolett.1c01661,PhysRevB.102.161109}, which has been reported for Fe in iron-based superconductors\cite{npjqm2_57}.}

\section{Summary}

In summary, our INS study on single crystals of Fe-deficient {\new\fgtn} has revealed that the spin excitation spectra are composed of a low-$E$ collective mode dispersing from the $\Gamma$ point, and a column-like continuum at the Brillouin zone boundary that extends to above 100~meV. We attribute these two modes to originate from ferromagnetic spin waves of the local moments and particle-hole continuum from the scattering between the spin-up and spin-down bands, respectively. The local spins and {\new some} itinerant electrons couple with each other through the Kondo effect, as evident from the unusual temperature effect of the low-$E$ spin-wave excitations. Our work provides a rare example that spin-wave excitations, particle-hole continuum and Kondo-lattice behavior coexist in a $d$-electron system, and will shed light on the understanding of magnetism in correlated electronic materials.

\section{Acknowledgements}

We would like to thank Prof.~Xingye~Lu at Beijing Normal University for allowing us to use their Laue machine. The work was supported by National Key Projects for Research and Development of China with Grant No.~2021YFA1400400, National Natural Science Foundation of China with Grant Nos~11822405, 12074174, 12074175, 11774152, 11904170, and 12004191, Natural Science Foundation of Jiangsu province with Grant Nos.~BK20180006, BK20190436 and BK20200738, {\newr Hubei Provincial Natural Science Foundation of China with Grant No.~2021CFB238}, Fundamental Research Funds for the Central Universities with Grant No.~020414380183, and the Office of International Cooperation and Exchanges of Nanjing University. We acknowledge the neutron beam time from J-PARC  with Proposal No.~2019A0002 and from ANSTO with Proposal No.~P8074. Z.M. thanks Beijing National Laboratory for Condensed Matter Physics for funding support.


%

\end{document}